%% file: radcor_proc.tex
\documentclass[12pt]{article}
\usepackage{epsfig}
\usepackage{wrapfig}

\newcommand\pubnumber{ANL-HEP-CP-01-10}
\newcommand\pubdate{\today}

\newcommand{\pom}{{\rm I\! P}}
\newcommand{\reg}{{\rm I\! R}}

\def\anl{Argonne National Laboratory\\
9700 S. Cass Ave, Argonne, Illinois 60439, USA}

\def\Title#1{\begin{center} {\Large #1 } \end{center}}
\def\Author#1{\begin{center}{ \sc #1} \end{center}}
\def\Address#1{\begin{center}{ \it #1} \end{center}}

\newcommand\pubblock{\rightline{\begin{tabular}{l} \pubnumber\\
         \pubdate  \end{tabular}}}
\newenvironment{Abstract}{\begin{quotation}  }{\end{quotation}}
\newenvironment{Presented}{\begin{quotation} \begin{center} 
             PRESENTED AT\end{center}\bigskip 
      \begin{center}\begin{large}}{\end{large}\end{center} \end{quotation}}
\def\Acknowledgements{\bigskip  \bigskip \begin{center} \begin{large}
             \bf ACKNOWLEDGEMENTS \end{large}\end{center}}

\input econfmacros.tex

\begin{document}
\begin{titlepage}
\pubblock

\vfill
\Title{HERA Small-$x$ and/or Diffraction}
\vfill
\Author{Rik Yoshida\\
{\small on behalf of the ZEUS and H1 Collaborations}}
\Address{\anl}
\vfill
\begin{Abstract}{
Recent HERA data on small-{\it x} structure
functions as well as DIS diffraction and diffractive
vector meson production are presented. The relationship
between these processes and possible indications
of dynamics beyond the DGLAP formalism are discussed.}
\end{Abstract}
\vfill
\begin{Presented}
5th International Symposium on Radiative Corrections \\ 
(RADCOR--2000) \\[4pt]
Carmel CA, USA, 11--15 September, 2000
\end{Presented}
\vfill
\end{titlepage}
\def\thefootnote{\fnsymbol{footnote}}
\setcounter{footnote}{0}
\section{Proton Structure Function $\boldmath F_2$ at Small-$\boldmath x$}

\subsection{Deep Inelastic Scattering and $\boldmath F_2$}
\begin{wrapfigure}{r}{2.3in}
\centerline{\epsfig{file=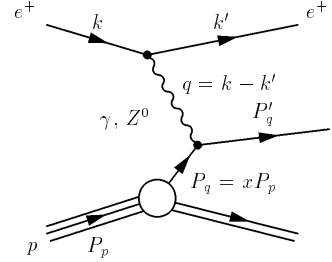,height=4.3cm,clip=}}
\caption{Collision of electron (positron), $e^\pm$, of four-momentum $k$ with
a proton, $p$, of four-momentum $P_p$.}
\end{wrapfigure}
Deep Inelastic Scattering (DIS) of electrons (or positrons) with a proton
is shown in Figure 1.  The reaction 
proceeds through the exchange of a virtual boson; in the kinematic range 
covered in this talk, only photon exchange is important.
The reaction can be described completely by two kinematic variables
chosen to be the four-momentum transfer squared, $Q^2=-q^2$ (see Figure 1), 
and the Bjorken variable, $x$.
In the Quark Parton Model, $x$ is the fraction of the initial
proton momentum carried by the struck parton.

The DIS cross-section factorizes into a short-distance
part which is the partonic cross-section, $\hat{\sigma}$, which can
be calculated perturbatively in QCD, and
a long-distance non-perturbative part, the parton densities, $f$.

\begin{wrapfigure}{l}{3in}
\centerline{\epsfig{file=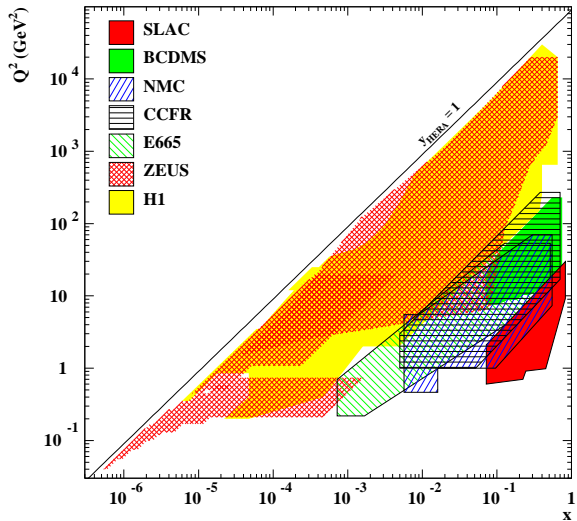,height=7.2cm,clip=}}
\caption{The kinematic region in which 
the proton $F_2$ has been measured. H1 and ZEUS are the experiments
at the HERA $ep$ collider.}
\vspace{0.4cm}
\end{wrapfigure}

At sufficiently high $Q^2$, the parton densities, $f$, obey 
the DGLAP equation\,\cite{dglap}, which 
is written schematically as,
\begin{equation}
{\frac{\partial f}{\partial \ln Q^2} \sim f\otimes P},
\end{equation}
where $P$ are the splitting functions that describe the branching
of quarks and gluons, and $\otimes$ symbolizes a convolution.

The DIS differential cross-section can be written in terms of 
the proton structure function $F_2$ as
\begin{equation}
{\frac{d \sigma^2}{dxdQ^2} = \frac{2\pi\alpha^2}{xQ^4}(1+(1-y)^2){ F_2(x,Q^2)}},
\end{equation}
where $y=Q^2/xs$ is the inelasticity parameter, and $s$ is 
the CMS energy squared of the $ep$ collision. The longitudinal 
structure function $F_L$ and the effects of $Z^0$ exchange have been neglected
in Equation (2).

\begin{wrapfigure}{l}{2.8in}
\centerline{\epsfig{file=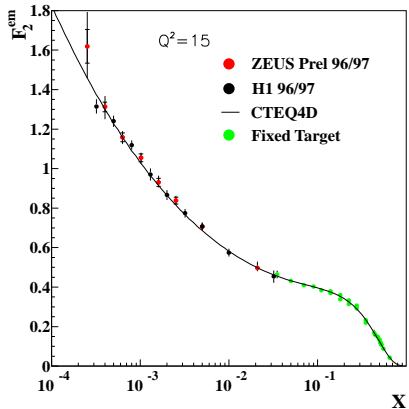,height=6cm,clip=}}
\caption{World's data on $F_2$ at $Q^2=15$ GeV$^2$ as a function
of $x$.  The solid line is a DGLAP fit by the CTEQ group \cite{cteq4}.}
\end{wrapfigure}

At leading order, 
\begin{equation}
{F_2(x,Q^2)} =x\sum_q e_q^2({q(x,Q^2)}+{\bar{q}(x,Q^2)}),
\end{equation}
where {$q$, $\bar{q}$} are the quark and antiquark distributions, respectively.

Figure 2 shows the $x$ and $Q^2$ range of the currently available
measurements of $F_2$.  At the HERA $ep$ collider, with $\sqrt{s}$ of 
about 300 GeV, $x$ and $Q^2$ can be varied over six orders 
of magnitude.  Of particular relevance to this talk is the region
of smallest $x$, which is probed only at HERA.

The measurements of $F_2$ at HERA show that the structure function
rises steeply at small $x$ (see Figure 3) \cite{h1f2,zeus9697}.  

If $F_2$ at $x <$ 0.1 is parameterized as $\propto x^{-\lambda}$,
then $\lambda$ falls as a function of $Q^2$ from about 0.4 to 0.1
as $Q^2$ falls from 200 GeV$^2$ to 1 GeV$^2$.
At $Q^2$ of 10 GeV$^2$, $\lambda$ is about 0.2 (Figure 4) \cite{h1lowq2,zeusphenom}.

\begin{wrapfigure}{r}{2.8in}
\centerline{\epsfig{file=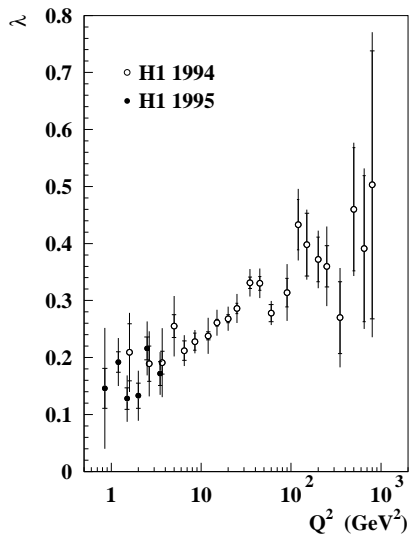,height=7cm,clip=}}
\caption{The parameter $\lambda$ of the $x$ dependence of $F_2$.  
See text and ref. \cite{h1lowq2}.}
\end{wrapfigure}

The naive physical interpretation of the small-$x$ rise of $F_2$ is
that it is caused by more and more gluons (and thus sea-quarks) being
present at smaller and smaller fractional momenta values, i.e. $x$.

The scaling violations of $F_2$ (i.e. the $Q^2$ dependence 
of $F_2$ at fixed $x$, as seen in Figure 5) at low $x$ 
are related, in Leading 
Order (LO) DGLAP, simply to the gluon density of the proton \cite{prytz},
\begin{equation}
\frac{\partial F_2(x/2,Q^2)}{\partial \ln Q^2} \propto {\alpha_s}xg(x,Q^2).
\end{equation}

In Next-to-LO (NLO) DGLAP, the simple relationship of Equation (4) no 
longer holds.  However,
the gluon density may be extracted from the NLO DGLAP fits to $F_2$. 

As an example of such fits, the one made by the H1 collaboration
is briefly described \cite{h1f2}.

The fit is made to the H1 data  
and the BCDMS $\mu p$ data at $Q^2 > 3.5$ GeV$^2$. The flavor 
decomposition of $F_2$ into 
the gluon $xg$, the valence component $V$, and the sea component
$A$ is done in such a way as to allow the use of proton data only
and avoid the deuteron data which introduces nuclear-correction
uncertainties.

\begin{wrapfigure}{l}{3in}
\centerline{\epsfig{file=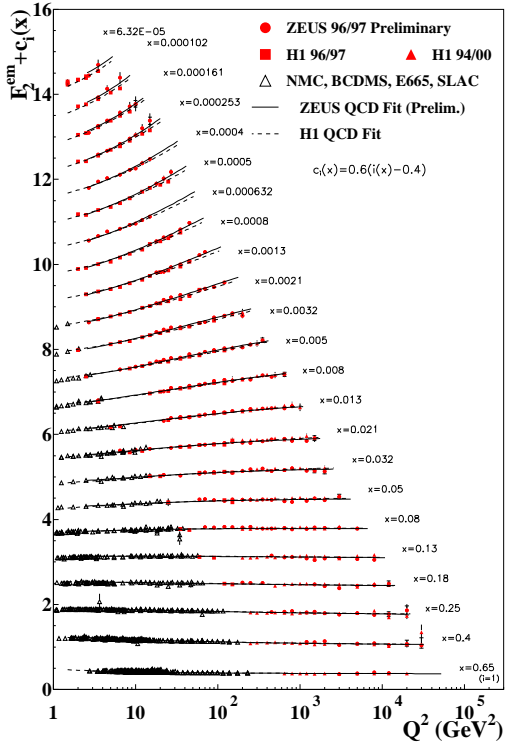,height=10cm,clip=}}
\caption{Proton $F_2$ for fixed $x$ as functions of $Q^2$.  A
constant, $c_i$, has been added to $F_2$ in order to make all of the
points visible.}
\vspace{-0.8cm}
\end{wrapfigure}

The $xg$, $V$, and $A$ distributions are parameterized in
the form $ax^b(1-x)^c \times$ $(1+d\sqrt{x}+ex)$.
The momentum sum rule is imposed leading to 16 parameters to
be fitted including $\alpha_s$.  
The fits are made by employing Equations (1) and (3) (in their NLO versions).

Figure 6 shows the gluon density extracted from the fit.  
In keeping with the naive expectation of a gluon-driven $F_2$,
the gluons also rise steeply at low $x$.  As $Q^2$ falls, the steepness
of the gluon also becomes less, as in the case of $F_2$ itself.

\subsection{DIS and $\boldmath \sigma_{tot}^{\gamma^*p}$}
Figure 2 shows that due to the kinematic limit at HERA, the measurements
of $F_2$ at the smallest $x$ values of $10^{-6}$--$10^{-5}$ correspond
to rather small values of $Q^2$, well below 1 GeV$^2$.  In this
kinematic range, it is appropriate to describe DIS in the
hadronic language of a collision between a virtual photon and
a proton.  The appropriate variable, in this case, becomes
the virtual-photon proton CMS energy, $W \approx \sqrt{Q^2/x}$.

The total virtual-photon proton cross-section, at small-$x$, can be
written in terms of $F_2$ as
\begin{equation}
{\sigma_{tot}^{\gamma^*p}(W^2,Q^2){= \frac{4\pi^2\alpha
}{Q^2}} F_2(x,Q^2)}.
\end{equation}
For fixed $Q^2$, $F_2 \propto x^{-\lambda}$ implies 
$\sigma_{tot}^{\gamma^*p} \propto W^{2\lambda}$.  The structure
function $F_2$ must vanish as $Q^2$, for fixed $W$, as $Q^2$
approaches 0, by conservation of EM current.
At $Q^2 = 0$ (photoproduction), the cross-section,
$\sigma_{tot}(W)$, is described 
by Regge phenomenology, and is known to agree 
with the universal hadron-hadron cross-section behavior at high
energies, $\sigma_{tot}(W) \propto W^{2(\alpha_\pom -1)}$ \cite{sigtot}.
The Pomeron intercept $\alpha_\pom$ has the value 1.08 \cite{dl}.
Thus at $Q^2 = 0$, $\sigma_{tot}(W) \propto W^{0.16}$, in contrast
to $\sigma_{tot}^{\gamma^*p}(W^2,Q^2) \propto W^{0.4}$ at $Q^2 \approx$
10 GeV$^2$.

\begin{wrapfigure}{r}{3in}
\centerline{\epsfig{file=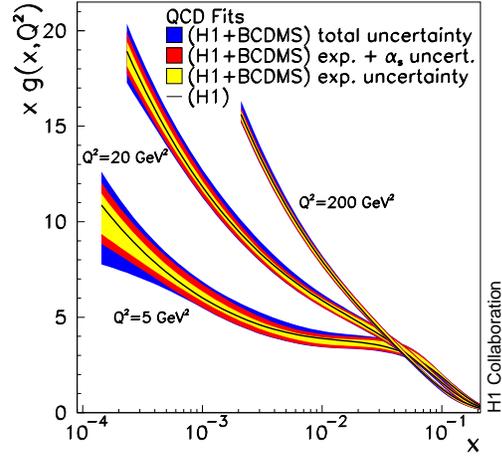,height=6cm,clip=}}
\caption{The gluon distribution extracted by the H1 Collab.  See
\cite{h1f2} for details.}
\vspace{-1cm}
\end{wrapfigure}

Figure 7 shows the measured $F_2$ at fixed $y = W^2/s$ 
down to the $Q^2$ value of
0.04 GeV$^2$.  $F_2$ begins to fall as $Q^2$ below
$Q^2 <$ 1 GeV$^2$.  The dashed line is a Regge inspired fit in the form 
$F_2(x,Q^2) = (\frac{Q^2}{4\pi^2\alpha})\cdot
(\frac{M_0^2}{M_0^2+Q^2})\cdot (A_\reg\cdot
(W^2)^{\alpha_\reg-1}+A_\pom\cdot(W^2)^{\alpha_\pom-1})$.
The $\alpha_\pom$ value of 1.1 gives a good fit. This is consistent
with the value of Pomeron intercept discussed above \cite{zeusphenom}.

\begin{wrapfigure}{l}{2.5in}
\centerline{\epsfig{file=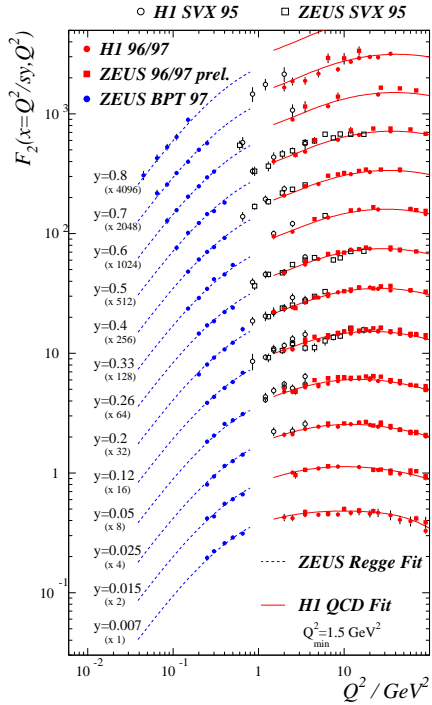,height=9.5cm,clip=}}
\caption{$F_2$ as a function of $Q^2$. The lines are Regge and pQCD fits.}
\end{wrapfigure}

Also shown in Figure 7 is a NLO DGLAP fit which 
gives a good description of the data from high $Q^2$ down to 
about 1 GeV$^2$.

\subsection{Beyond DGLAP?}

The interest in small-$x$ physics is that the partons under study
are the result of a large number of QCD branching processes.
The evolution of the number of partons over a wide kinematic range
in $x$ and $Q^2$ should be sensitive to the applicability
of different perturbative approximations of QCD.

Figure 8 shows the qualitative expectation of applicability of
various pQCD approaches.  DGLAP is a resummation of terms
proportional to ($\ln{Q^2}$) and is expected to hold in the region
of large $Q^2$.  BFKL \cite{bfkl} is a resummation of terms proportional
to ($\ln{1/x}$) and, while the stability of the perturbative
expansion still under study \cite{bfklnlo,blmln}, it is expected to hold in 
the region of small $x$. The CCFM \cite{ccfm}
equation incorporates both ($\ln{1/x}$) and ($\ln{Q^2}$) terms.

\begin{wrapfigure}{r}{2.9in}
\centerline{\epsfig{file=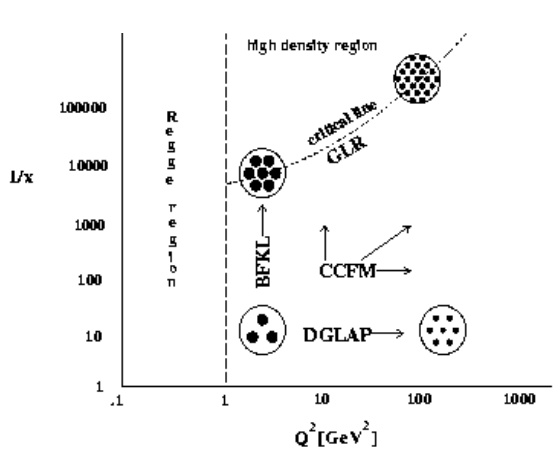,height=5.5cm,clip=}}
\caption{Schematic diagram of the applicability of different
pQCD approximations.}
\end{wrapfigure}

At small enough $x$, the density of partons should become sufficiently
large so that the interactions between them become important.  This
boundary is marked in Figure 8 by the line labeled ``critical line".
The GLR equations \cite{glr,blmln2} attempt to take these saturation, 
or shadowing, effects into account.

It has been shown above that the DGLAP formalism is able to 
describe the currently available $F_2$ data down to 1 GeV$^2$
and $x$ of $10^{-5}$ in apparent contradiction to Figure 8,
at least with the ${1/x}$ scale numbers as drawn.

On the other hand, there have also been successful fits to a wide
range of $F_2$ data using formalisms that incorporate the ($\ln{1/x}$) terms
as well as the ($\ln{Q^2}$) terms \cite{thorne,jung}.

In search of clarification, 
we turn next to the phenomenon of DIS diffraction, and vector
meson production, before returning to consider if there are
any indications in the $F_2$ data for dynamics beyond DGLAP.

\section{Diffraction in DIS}

One of the striking results from HERA is the presence of
diffractive events in DIS \cite{zeusdiff,h1diff}.  
About 10\% of all DIS events have
a gap in particle emission 
between the final-state proton, or a low mass state,
which travels down the beampipe, and the system $X$, which is
measured in the detector (Figure 9).
Such a reaction is usually described as an exchange of 
a colorless object, generically called the Pomeron ($\pom$).

\begin{wrapfigure}{r}{2.9in}
\centerline{\epsfig{file=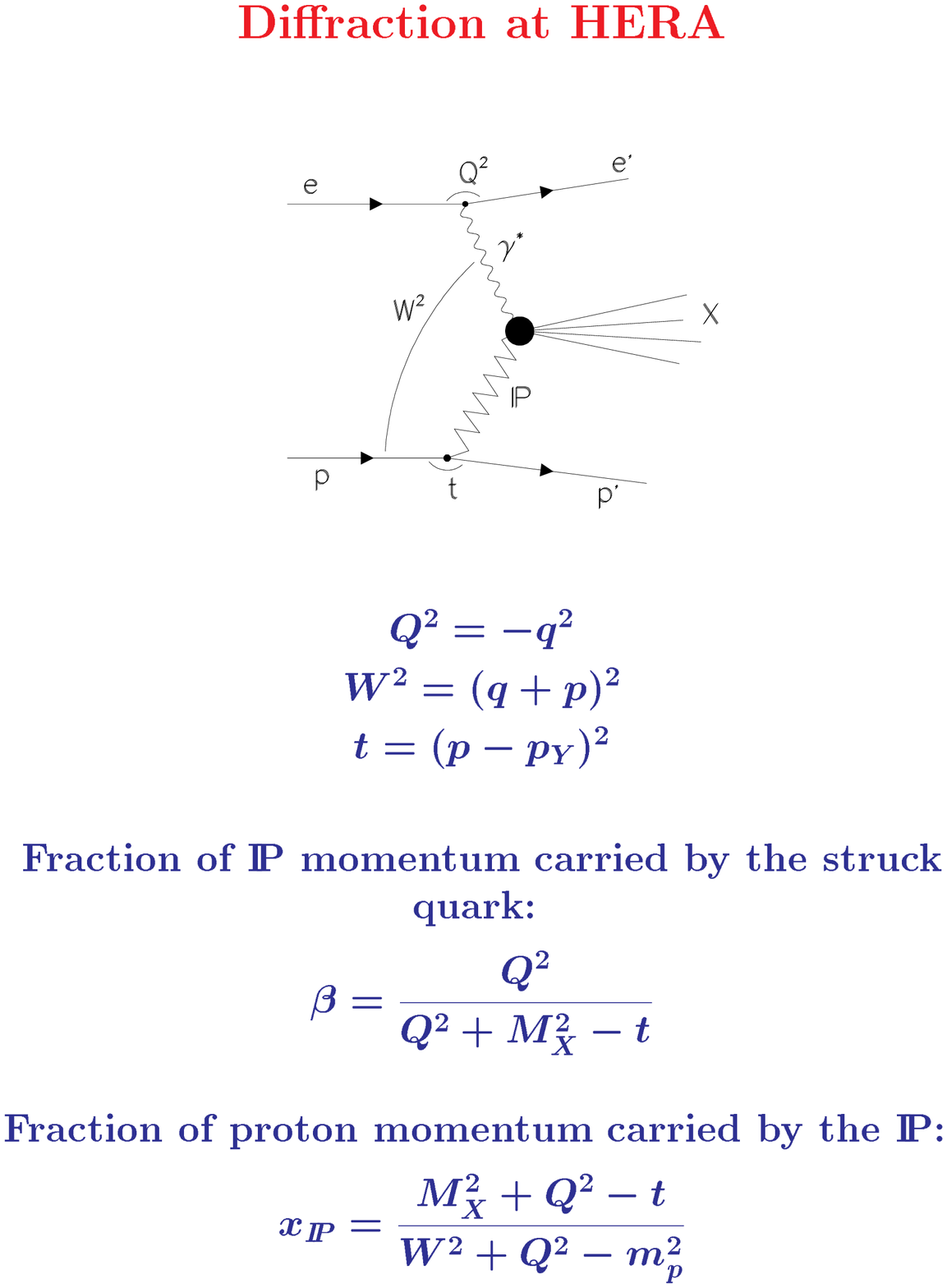,height=6cm,clip=}}
\caption{Diffractive DIS scattering.}
\vspace{-0.1cm}
\end{wrapfigure}
 
In order to describe diffractive DIS, two kinematic variables in addition
to $x$ and $Q^2$ are needed.  These are $t$, which is the momentum
transfer at the proton vertex and $x_\pom$, which is the fractional
momentum of the proton carried by the Pomeron.  Another useful
variable is $\beta=x/x_\pom$, which has an interpretation as the
fractional momentum of the Pomeron carried by the struck parton 
(i.e. the Pomeron analogue of $x$ for the proton).

\begin{wrapfigure}{r}{2.9in}
\centerline{\epsfig{file=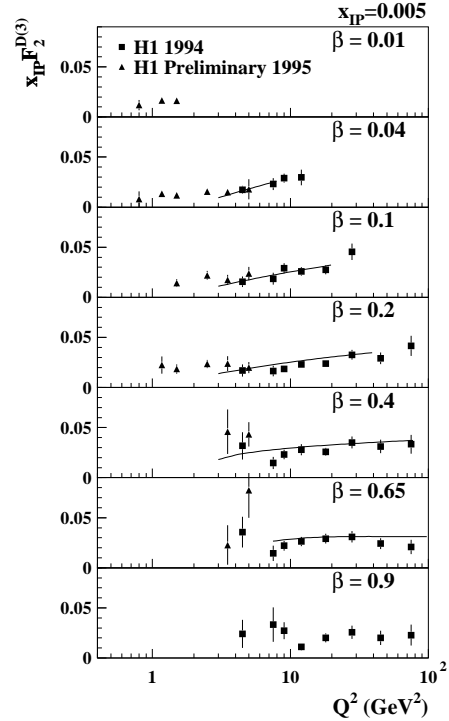,height=10cm,clip=}}
\caption{The diffractive structure function for a bin of $x_\pom$ measured
by the H1 Collab. See text.
}
\vspace{-0.5cm}
\end{wrapfigure}

Any perturbative description of diffractive DIS
must go beyond the simplest DGLAP picture; the lack of color connections
between the system $X$ and the proton must mean that at least
two gluons are exchanged.  It is then interesting to investigate
the connection between the high gluon densities implied by the $F_2$
measurements and the phenomenon of DIS diffraction.

\subsection{Diffractive Factorization and Pomeron Structure}

The diffractive cross-section, in analogy with the total cross-section,
is written in terms of the diffractive structure function $F_2^D$
as
\begin{eqnarray}
\frac{d^3\sigma^D}{d\beta dQ^2 dx_{\pom}} =
  \frac{2\pi\alpha^2}{\beta Q^4}(1+(1-y)^2) \nonumber\\
 \times F_2^{D(3)}(\beta,Q^2,x_{\pom}).
\end{eqnarray}
The cross-section has been integrated over $t$.  It has been proven
that $F_2^D$ factorizes into a long and a short distance contributions,
as does the inclusive $F_2$,
i.e. $F_2^D\sim f^{D} \otimes \hat{\sigma}$
where $\hat{\sigma}$ are the usual pQCD hard cross-sections and
$f^D$ are the diffractive parton densities, which obey the usual
DGLAP equations, and are universal \cite{collins}.  By knowing $f^D$, we can
calculate any diffractive DIS final state such as charm or jet production.

The diffractive parton densities are functions of four variables:
$x_\pom$, $t$, $\beta$ and $Q^2$.  However, the DGLAP evolution
only concerns the variables $x$ (or $\beta$) and $Q^2$.  If, for
all relevant $f^D$'s, the $x_\pom$ and $t$ dependences decouple
from the $\beta$ and $Q^2$ dependence, {\it and} if the
$x_\pom$ and $t$ dependences are the {\it same} for all relevant
partons, then we arrive at what is known as Regge factorization \cite{is},
\begin{equation}
F_2^D(x_{\pom},t,Q^2,\beta) =f(x_\pom,t)\cdot F_2^{\pom}(\beta,Q^2).
\end{equation}
In this case $f(x_\pom,t)$ can be interpreted as the flux
factor of the Pomeron, and $F_2^{\pom}$ as the structure function
of the Pomeron.  In this case DGLAP analysis of $F_2^{\pom}$ becomes
meaningful.
It is a remarkable experimental fact that Regge factorization,
which is not required by the diffractive factorization theorem,
apparently holds over a large part of the measured phase-space,
and that the flux factor, $f(x_\pom,t)$, has 
approximately the form expected by Regge theory of 
$1/x_\pom^{2\bar{\alpha}_\pom-1}$ \cite{zeusdiff,h1diff} (see also Figure 17a).

\begin{wrapfigure}{l}{2.5in}
\centerline{\epsfig{file=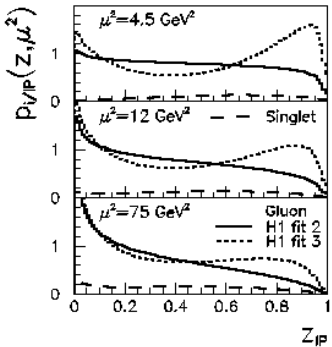,height=5cm}}
\caption{The parton distribution, $P$, of the Pomeron extracted by
the H1 Collab.  See text.}
\end{wrapfigure}

Figure 10 shows the measurements of $F_2^{D(3)}$ by the H1 collaboration
for $x_\pom$ of 0.005 \cite{h1diff,h1diff2}.  
The lines are the results of the DGLAP analysis.
The resulting parton distributions in the Pomeron is shown in Figure 11.

The analysis finds two stable solutions, one of which (dotted line)
favors a rather large amount of gluons at $z_{\pom}=1$, and
the second which does not (solid line).  The fractional momentum 
carried by the parton in the Pomeron is denoted by $z_{\pom}$, and the 
renormalization and factorization scale by $\mu$.

The recent measurement by H1 of dijet production in 
diffraction \cite{h1diffjets} shows
that the Pomeron parton distributions extracted from the DGLAP analysis
can indeed be used to describe the dijet cross-section in diffractive
events (Figure 12).
The jet measurement favors the parton distributions labelled ``fit 2"
in Figure 11.

\subsection{Diffraction and $F_2$ at Small-$x$ }

\begin{wrapfigure}{r}{2.5in}
\centerline{\epsfig{file=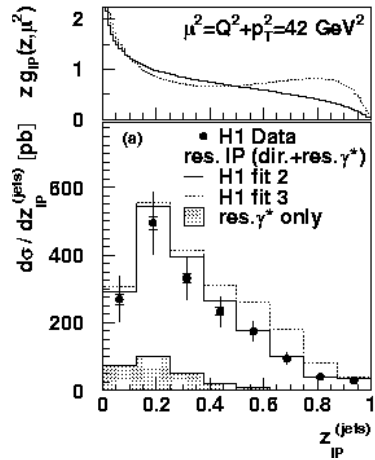,height=6cm,clip=}}
\caption{Dijet cross-section for diffractive events compared to predictions
based on the gluon momentum distribution of the Pomeron, $zg_\pom$. See text
and ref \cite{h1diffjets}.}
\end{wrapfigure}

While the analyses based on diffractive factorization have been very
successful and powerful, the question of the origin of the 
phenomenon of DIS diffraction remains unanswered.  Furthermore, the
relationship between the Pomeron structure and the proton structure
is not clear.

Figure 13 shows the ratio, for fixed $Q^2$, of the DIS diffractive cross 
section to the total DIS cross-section as 
measured by the ZEUS collaboration.
Although a cut has been  made in the mass of the diffractive system,
$M_X$, the conclusion is independent of $M_X$: the ratio
is flat as a function of $W$ or, 
equivalently for fixed $Q^2$, of $x$ \cite{zeusdiff}.
The flatness of the ratio implies that the energy dependence of
DIS diffraction is the same as that of inclusive DIS, i.e. $\propto
W^{0.4}$ at $Q^2 \approx 10$ GeV$^2$.

\begin{wrapfigure}{r}{3.5in}
\centerline{\epsfig{file=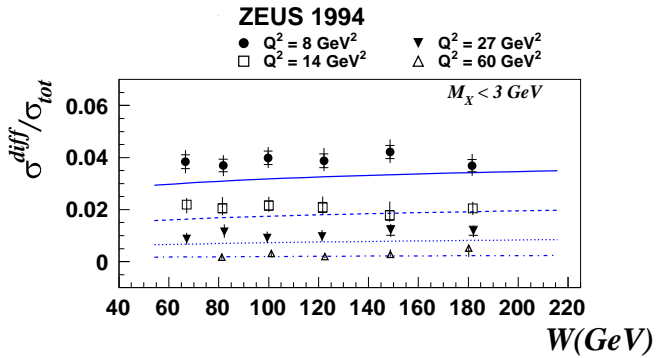,height=5cm,clip=}}
\caption{The ratio of diffractive to total DIS cross-section. The
data have been binned in $Q^2$ and a cut on the mass of the
state $X$, $M_X <$ 3 GeV has been made.}
\vspace{-1cm}
\end{wrapfigure}

This result is surprising from several points of view.  A naive expectation
from the optical theorem would lead to diffractive cross 
sections
that rise twice as fast as the inclusive one.  In other words,
if the rise of the inclusive cross-section is driven by a gluon density
that rises as $x^{-\lambda}$ (or $W^{2\lambda}$), then for diffractive
process that needs to couple to at least two gluons, the cross-section 
should rise as $x^{-2\lambda}$.
At the same time, the energy dependence of the diffractive cross-section
also contradicts Regge phenomenology, which expects
an energy dependence of $W^{0.25-0.3}$.

The lines in Figure 13 that describe the data qualitatively are from
the dipole model of Golec-Biernat and W\"{u}sthoff 
\cite{gbw1,gbw2}, briefly
described below.

\subsection{Impact Parameter Space (or Dipole) Models}

The infinite momentum frame of DIS, which is appropriate
for the DGLAP formalism, is obviously not the only possible frame of
reference.  In the impact parameter, or dipole, formalism, the
appropriate frame of reference is that in which the 
virtual photon dissociates 
into a quark-anti quark pair (or a more complicated state), which 
forms a color dipole which collides with the proton (Figure 14).

\begin{wrapfigure}{l}{2in}
\centerline{\epsfig{file=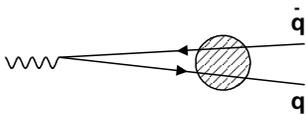,height=2cm,clip=}}
\caption{Color dipole picture of DIS}
\end{wrapfigure}

The description of the interaction of this color dipole with 
the proton distinguishes the various models of this type that 
have been proposed \cite{dipoles}.
In this presentation, a particularly simple model due to 
Golec-Biernat and W\"{u}sthoff (GB\&W) will be described.

In the GB\&W model, the cross-section of the dipole, $\hat{\sigma}_{dipole}$,
with the proton
is modeled simply as a function that smoothly interpolates between
two limits (Figure 15); 
at small dipole radius, $r$, $\hat{\sigma}_{dipole}$ 
increases as $r^2$ in keeping with the 
behavior of perturbative QCD.  At large $r$, $\hat{\sigma}_{dipole}$
becomes constant to preserve unitarity.  The saturation of the dipole
cross-section can be qualitatively shown to correspond to 
the saturation of partons in the proton \cite{mueller}.

In the model, the point at which the $r^2$ dependence of the
cross-section changes to the
constant behavior depends on the density of the partons in the
proton. Specifically, a parameter, $R_0 \sim 1/xg(x) \sim x^\lambda$,
which can be interpreted as the separation of partons
in the proton, is introduced.  When $r \ll R_0$, $\hat{\sigma}_{dipole}$
is in the ``pQCD" region, while when $r \gg R_0$, it is in the
saturation region (Figure 16a and 16b, respectively).

\begin{wrapfigure}{l}{4in}
\centerline{\epsfig{file=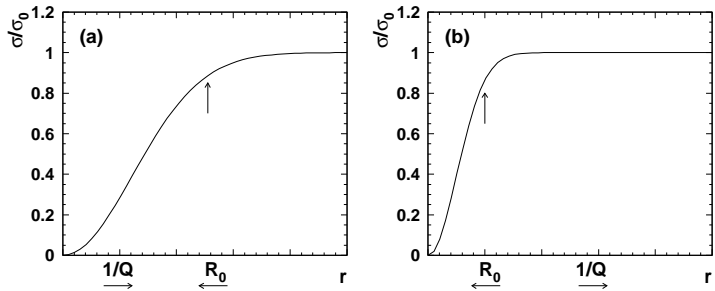,height=4cm,clip=}}
\caption{The dipole cross-section of the GB\&W model for (a) the
case of small dipole radius and (b) large dipole radius.}
\vspace{-1.2cm}
\end{wrapfigure}

The dipole radius, $r$, enters through the wave function of
the virtual photon, $\Psi_\gamma$, and the diffractive cross-section
is written, up to a $t$ dependence, as,
\begin{eqnarray}
\sigma^{D}\propto \int d^2r \int dz \nonumber\\
\times |\Psi_\gamma(z,r)|^2\hat{\sigma}_{dipole}^2(x,r),\nonumber\\
(8)\nonumber 
\end{eqnarray}
where $z$ is the fractional momentum of one of the quarks in the dipole.
\setcounter{equation}{8}

The diffractive cross-section is explicitly related to
the total cross-section, which is,
\begin{equation}
\sigma_{tot}\propto \int d^2r \int dz |\Psi_\gamma(z,r)|^2\hat{\sigma}_{dipole}(x,r).
\end{equation}
The fact that $\sigma^{D}/\sigma_{tot}$ is constant as a function 
of $W$ (or $x$), as shown in Figure 13, can be shown to occur
only if the dipole cross-section is being probed in
these processes beyond the small $r$ region into the saturation 
region.  The implication of this for the total cross-section
will be discussed below.

\begin{wrapfigure}{r}{4in}
\centerline{\epsfig{file=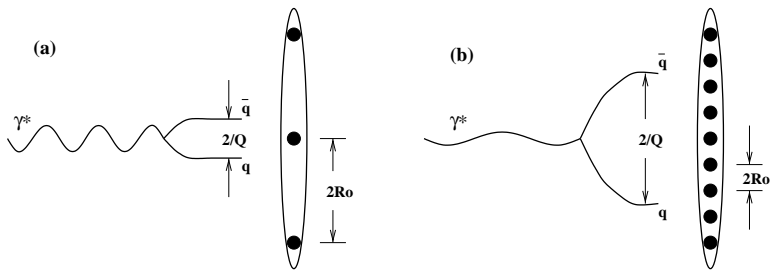,height=3.5cm,clip=}}
\caption{Schematic representation of the GB\&W model.  The characteristic 
radius of the dipole is proportional to the $Q^2$ of the 
virtual photon, $\gamma^*$.
The parameter $R_0$ corresponds to the separation of the partons
within the proton.  The relative size of $r$ and $R_0$ determines
the behavior of the dipole cross-section.
}
\vspace{-1.5cm}
\end{wrapfigure}

In Figure 17a, the measurements of $F_2^{D}$, this time from the
ZEUS collaboration, are shown along with the prediction of the GB\&W 
model.  There is qualitative agreement.  
Figure 17b shows 
the recent measurement of $F_2^{D}$ at very low $Q^2$ \cite{zeusdifflow}.  
The
GB\&W model again describes the data
qualitatively.

\newpage

\begin{figure}
\centerline{\epsfig{file=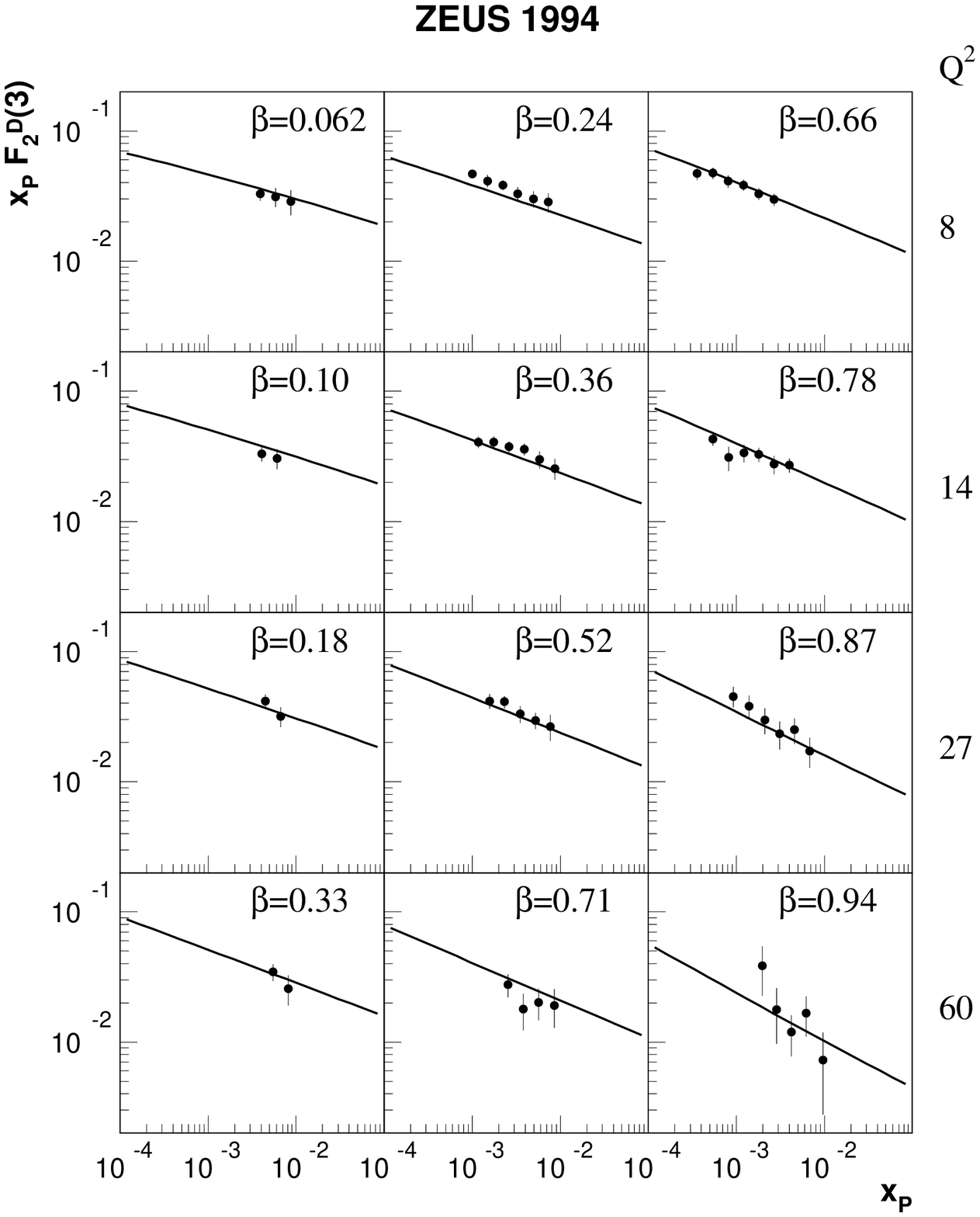,height=8.0cm,clip=}\epsfig{file=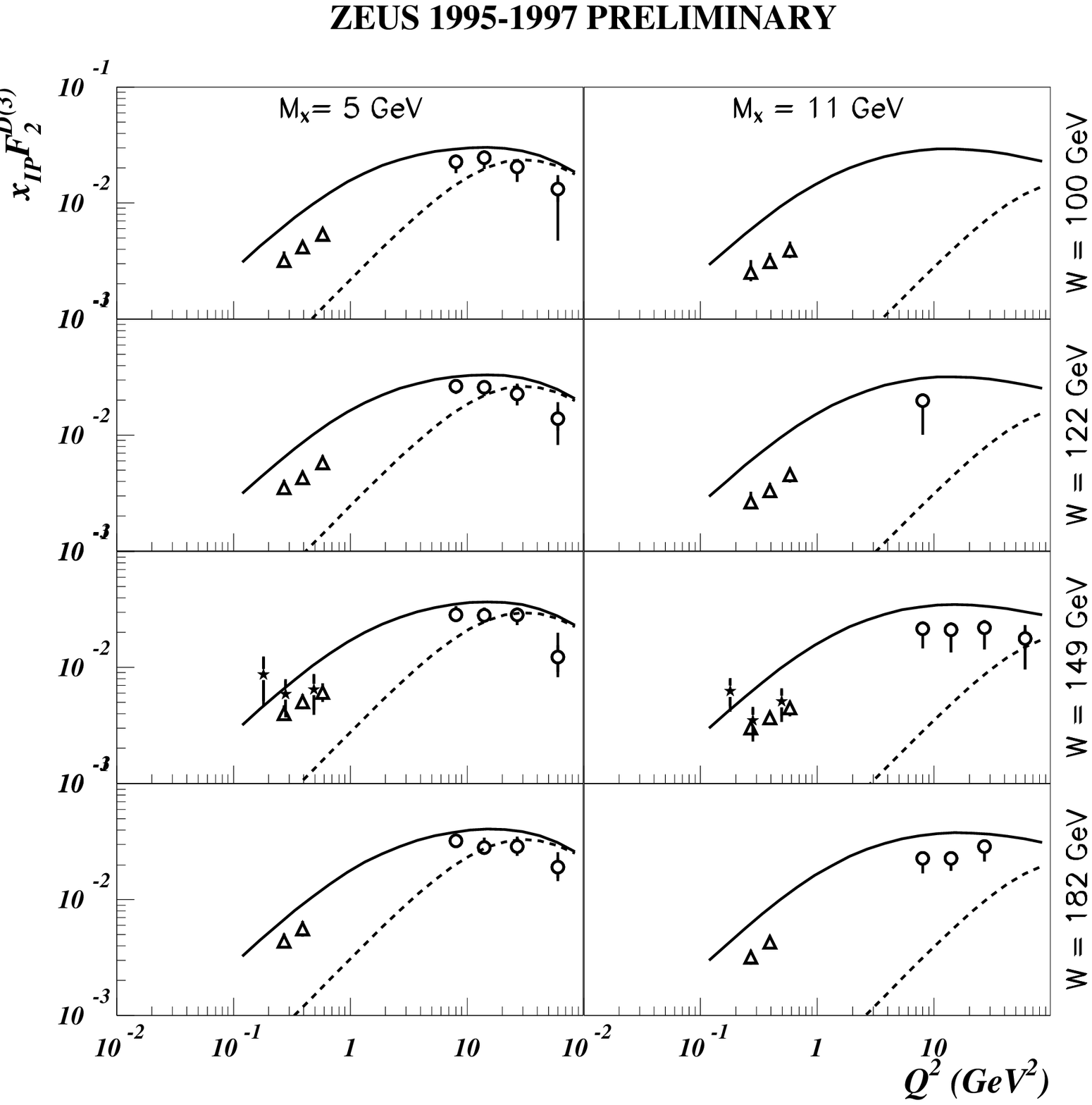,height=8.0cm,clip=}}
\caption{GB\&W model (solid lines) compared
to $x_\pom F_2^D(3)$ measurements. 
See text and refs \cite{zeusdiff,gbw2,zeusdifflow}.}
\begin{picture}(0.5,0.5)
\put(50,268){(a)}
\put(235,268){(b)}
\end{picture}
\end{figure}

\section{Diffractive Vector Meson Production}

\begin{wrapfigure}{l}{3in}
\centerline{\epsfig{file=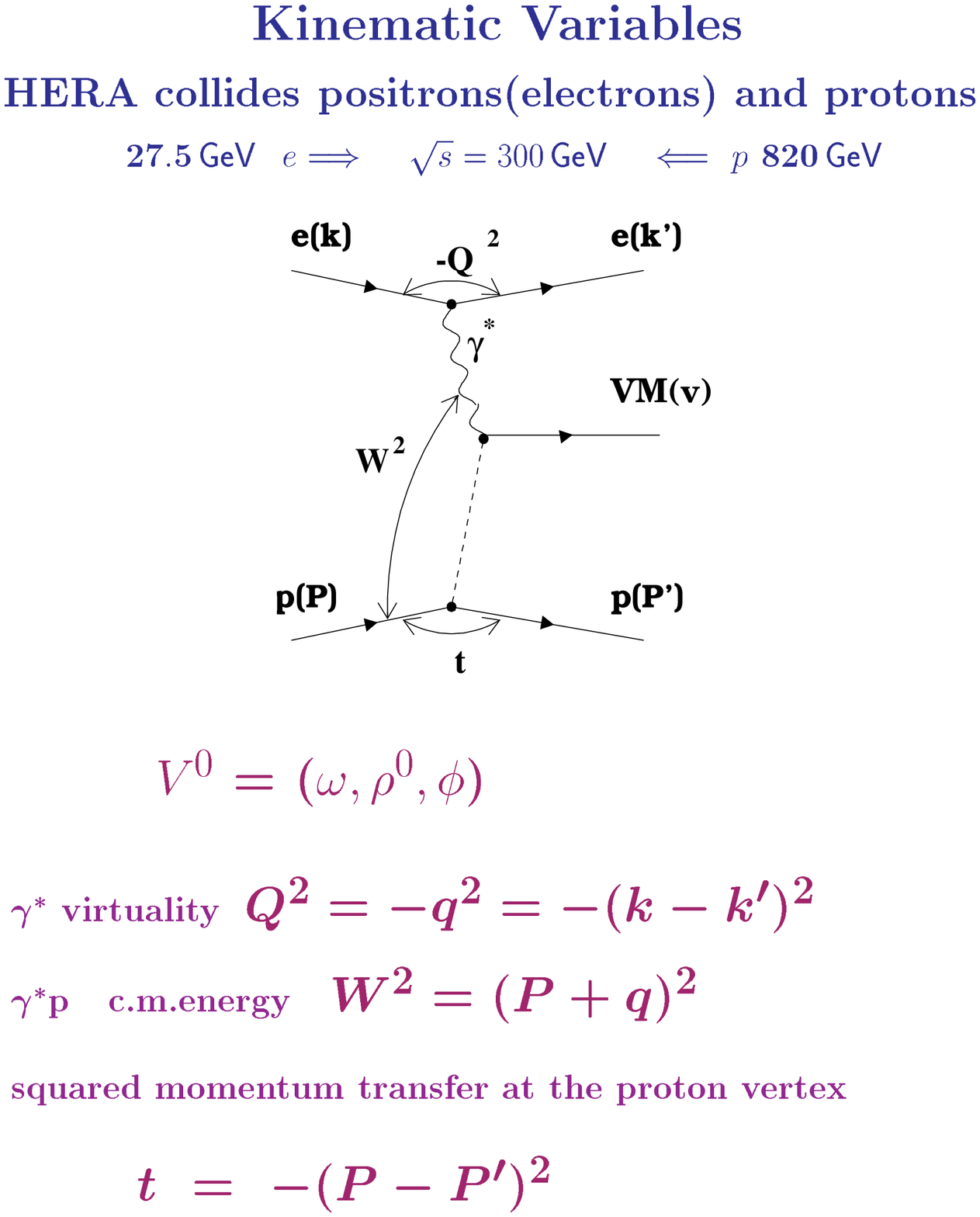,height=4cm,clip=}}
\caption{Diffractive VM production.}
\vspace{-0.6cm}
\end{wrapfigure}

The exclusive production of vector mesons (VM) in DIS is a process
which is closely related  to DIS inclusive diffraction.  As seen in Figure
18, the final state X in Figure 9 is replaced by a vector meson.
Let us review the energy dependences of the processes we have 
discussed so far:

\begin{itemize}
\item{The total real photoproduction cross-section, $\sigma_{tot}^{\gamma p}$,
increases in accordance with the Regge expectation, i.e. $W^{2(\alpha_\pom-1)}
\approx W^{0.16}$.}
\item{The total virtual-photon proton cross-section,
$\sigma_{tot}^{\gamma^* p}$, at $Q^2 \approx 10$ GeV$^2$ increases
approximately as $W^{0.4}$, corresponding to $F_2$ rising as $x^{-0.2}$.
}
\item{The diffractive DIS cross-section, $\sigma^{D}$, 
has approximately the same energy dependence as $\sigma_{tot}^{\gamma^* p}$,
i.e. increases as $W^{0.4}$ at $Q^2 \approx 10$ GeV$^2$.} 
\end{itemize}

\begin{wrapfigure}{r}{3.2in}
\centerline{\epsfig{file=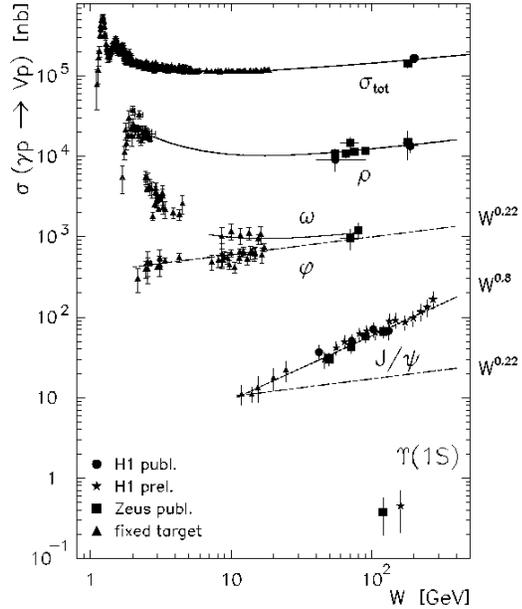,height=8cm,clip=}}
\caption{The $W$ dependence of diffractive VM photoproduction.}
\end{wrapfigure}

The energy dependence of exclusive vector meson production at $Q^2=0$
(photoproduction) is summarized in Figure 19.  While the production
of light vector mesons, $\rho$, $\phi$ and $\omega$, have a weak
energy dependence consistent with Regge expectations, the $J/\psi$
production cross-section rises rapidly with energy, as
$W^{0.8}$, i.e. twice as 
fast as $\sigma_{tot}^{\gamma^* p}$, at 
$Q^2 \approx 10$ GeV$^2$. It is the fastest energy 
dependence of the processes discussed so far \cite{h1jpsi,zeusjpsi}. 

The $Q^2$ dependence of the $J/\psi$ production cross-section has
been studied, but is still statistics limited.  There is no evidence
for a change of the energy 
dependence up to $Q^2$ of 15 GeV$^2$ \cite{zeusjpsi2}.

The $Q^2$ dependence of $\rho$ production cross-section is shown
in Figure 20 \cite{zeusrho}, which shows $\delta$, as obtained
from a fit to the form $W^\delta$, as a function of $Q^2$. 
The energy dependence clearly rises as a function of $Q^2$, although
the precise form is not yet clear from the limited statistics.

\begin{wrapfigure}{l}{2.8in}
\centerline{\epsfig{file=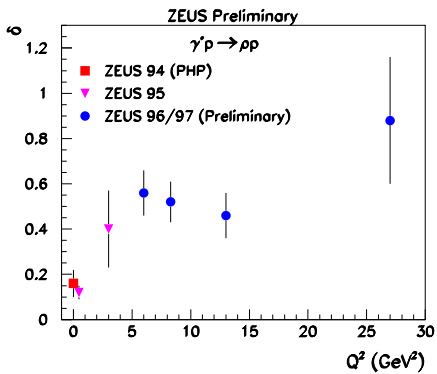,height=6cm,clip=}}
\caption{The energy dependece of $\rho$ production as a function
of $Q^2$.}
\end{wrapfigure}

In Figure 21, $\delta$ for various species of vector mesons has
been plotted as a function of $Q^2 + M_{VM}^2$, where
$M_{VM}$ is the mass of the vector meson.  While the statistics
are still limited, the data are consistent with a scaling of $\delta$ 
with $Q^2 + M_{VM}^2$.  If we recall that 
$\delta=2\lambda$ (see Section 1.2), we can compare the energy
dependence of vector meson production with that of inclusive
DIS.

Comparing Figure 21 with Figure 4 
which shows $\lambda$ extracted from fits to $F_2$ to the
form $x^{-\lambda}$ for $x < 0.1$, it can be seen that
$\lambda_{(VM)}(Q^2 + M_{VM}^2) \approx 2\lambda_{(F_2)}(Q^2)$.  This means
that the vector meson production has an energy dependence at
$Q^2 + M_{VM}^2$ which is twice as fast as that of the inclusive
DIS cross-section at $Q^2$.

\begin{wrapfigure}{r}{3.8in}
\centerline{\epsfig{file=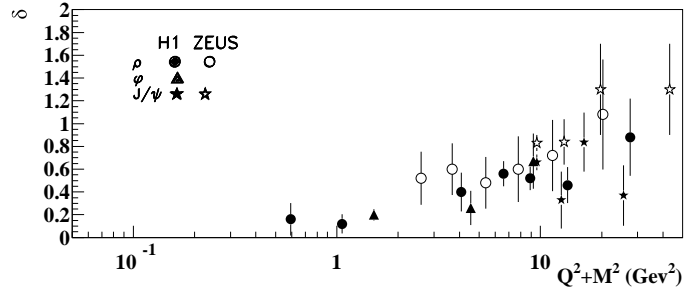,height=4cm,clip=}}
\caption{Energy dependence parameter $\delta$ for different VM's plotted
as a function of $Q^2 + M_{VM}^2$.}
\end{wrapfigure}

The ratio of cross-sections of different species of vector mesons
to the $\rho$ meson is shown in Figure 22 as a function of $Q^2 + M_{VM}^2$.
The ratio is flat as a function of $Q^2 + M_{VM}^2$ and is consistent
with the ratios $\rho:\omega:\phi:J/\psi$ being 9:1:2:8. Such ratios
are expected from the coupling of a photon
to a vector meson, $f^2_{VM}$, from 
simple charge counting of the quark content of the vector mesons.

\begin{wrapfigure}{l}{3.2in}
\centerline{\epsfig{file=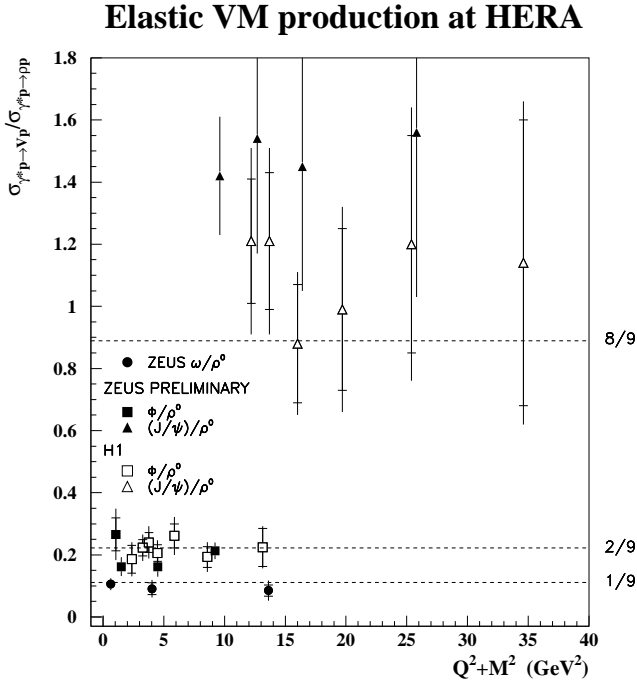,height=9cm,clip=}}
\caption{Ratios of diffractive VM production cross-sections to 
diffractive $\rho$ production cross-section, as a function 
of $Q^2 + M_{VM}^2$. 
}
\end{wrapfigure}

In the Vector Meson Dominance Model (VDM) of the production of 
vector mesons in photon 
(or virtual photon) proton collisions, the cross-section is
given as $\sigma_{VM}=f_{VM}^2 \cdot \sigma_{VMp}$, where
$\sigma_{VMp}$ is the vector-meson proton cross-section.
The fact that the ratio of the cross-sections is given by $f_{VM}^2$ implies
that $\sigma_{VMp}$ for different species of vector mesons scales
as $Q^2 + M_{VM}^2$.  In the context of color dipole models, the
implication is that $Q^2 + M_{VM}^2$ gives a measure of the 
radius, $r$, of the color dipole, which is the primary variable
on which the cross-section depends.

Figure 23 shows the values of the $t$ slope, $b$, where the vector meson
cross-sections are parameterized as $d\sigma_{VM}/d|t| \propto e^{-b|t|}$.
The parameter $b$ is plotted as a function of $Q^2 + M_{VM}^2$.
Within the statistical accuracy of the measurements,
$b$ scales with $Q^2 + M_{VM}^2$ and decreases
with increasing $Q^2 + M_{VM}^2$.    This is an indication that
$Q^2 + M_{VM}^2$ is indeed giving the effective radius of the interaction,
and thus the radius of the vector meson, or the color dipole.

\begin{wrapfigure}{l}{3.8in}
\centerline{\epsfig{file=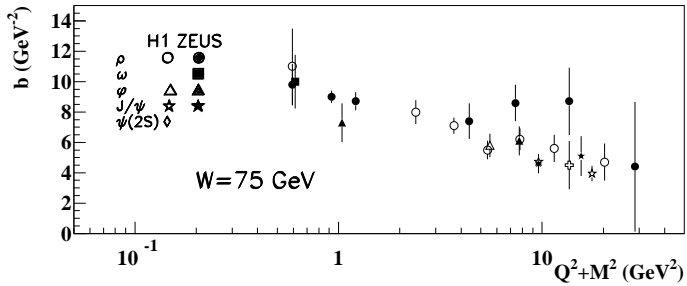,height=4cm,clip=}}
\caption{The slope dependence parameter $b$ for different VM's plotted
as a function of $Q^2 + M_{VM}^2$.}
\vspace{-0.5cm}
\end{wrapfigure}

The cross-section for diff\-ractive
vector meson production can be written in the color dipole model as,
\begin{eqnarray}
\sigma_{VM} \sim |\int dz \int d^2r \Psi_{VM}  \nonumber\\
\times {\hat{\sigma}_{dipole}} \Psi_{\gamma}|^2,
\end{eqnarray}
where $\Psi_{VM}$ is the wave function of the vector meson, which,
in contrast to the virtual photon wave function, $\Psi_{\gamma}$, is not
a well known quantity.  Recently, evaluation of 
vector meson production in the context of 
Equation (10) and the dipole cross-section,
$\hat{\sigma}_{dipole}$, from the GB\&W model has been made \cite{caldwell}.
Simple wave-functions
for the vector mesons were chosen.  The results reproduce the 
elastic $J/\psi$ cross-section and their energy dependence very well.
For the lighter mesons, $\phi$ and $\rho$, 
the calculations fail to give a good
description of the magnitude of the cross-sections. However, 
the steepening of the energy dependence with increasing $Q^2$ (Figure 20) 
is well reproduced.

\section{$\boldmath F_2$ at Small-$\boldmath x$ Revisited}

In the first part of this presentation, it was stated that the 
proton $F_2$ measured so far at HERA and elsewhere is well described
by fits using DGLAP equations above $Q^2$ of 1 GeV$^2$, and a model
combining Regge theory and Vector Meson Dominance has been shown to
work well below 1 GeV$^2$.

The dipole model description of the total DIS cross-section (or $F_2$)
is given in  Equation (9).  Currently the formal connection between the
DGLAP interpretation and the dipole models is far from clear \cite{mueller,
bartels,mcdermott}.
However many dipole models, in particular the GB\&W model,
describe the dynamics of cross-section saturation  
which goes beyond the physics
of the DGLAP picture. Thus, it is appropriate to revisit
the inclusive DIS data and ask if there are any indications of behavior
beyond DGLAP, even though the DGLAP fits are successful.

The DGLAP evolution equations, at low $x$ and at LO, 
imply that the partial derivative of $F_2$ with respect to ($\ln{Q^2}$)
is proportional to $xg(x,Q^2)$, the gluon momentum 
density of the proton (Equation (4)). 
On the other hand, 
at sufficiently low $Q^2$, $F_2$ must
vanish as $Q^2$ from current conservation, a behavior built into
the GB\&W model but not in the DGLAP formulation of DIS:
\begin{equation}
\frac{\partial F_2}{\partial \log{Q^2}} \propto Q^2\sigma_0.
\end{equation}

\begin{wrapfigure}{l}{4in}
\vspace{-1.0cm}
\centerline{\epsfig{file=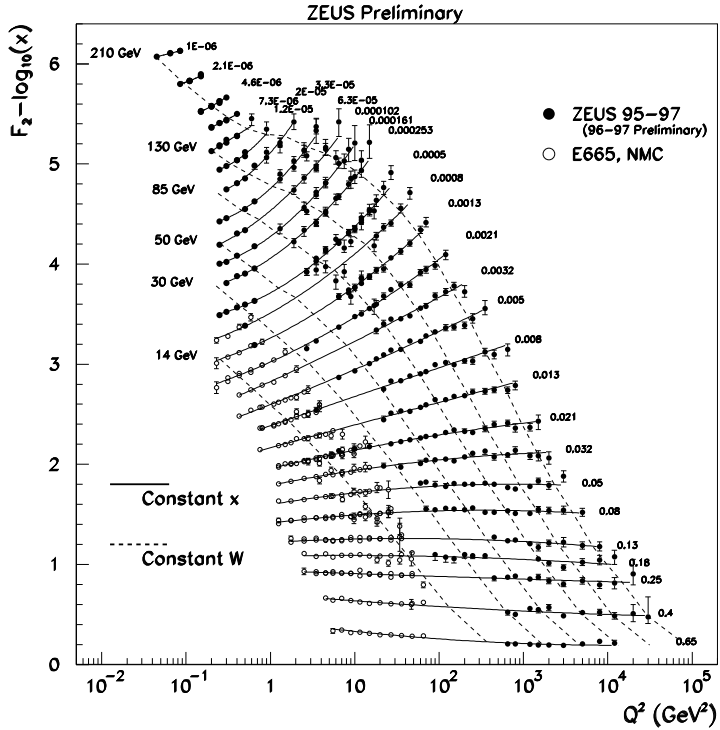,height=11cm,clip=}}
\caption{$F_2$ measurements in bins of $x$ as a function of $Q^2$.  
The solid lines
are the results of the fits described in the text.}
\vspace{0.3cm}
\end{wrapfigure}

Since $W$ at low $x$ is related to
$x$ and $Q^2$ by $W^2\approx Q^2/x$,
the plot of ${\partial F_2}/{\partial \log{Q^2}}$ 
at a constant $W$, over the $Q^2$ range covered by the data, should 
show a transition from the behavior according to  Equation (4) to
that of  Equation (11), provided $xg(x,Q^2)$ has a weak $Q^2$ dependence 
and $\sigma_0$, which corresponds to the photoproduction
cross-section, has a weak $x$ (or equivalently energy) dependence.  
Since there is no quantitative prediction for the kinematic
region in which the DGLAP formalism is applicable, the 
determination of the $Q^2$ and $x$ at which this transition occurs 
is of great interest and may help in clarifying 
the nature of the transition \cite{zeusslopes}.

Figure 24 shows recent measurements of $F_2$ from 
the ZEUS collaboration
along with the fixed target measurements~\cite{e665nmc} 
from the NMC and the E665 collaborations.
$F_2$ is shown as a function of $Q^2$ for bins of fixed $x$.  Where 
necessary, the measurements have been interpolated to the appropriate values
of $x$ using the ALLM parameterization~\cite{allm}.
The $F_2$ in each bin of fixed $x$ is offset by an 
additive factor of $(-\log_{10}x)$ to ensure that the 
vertical separation between the bins is monotonic in $x$.
Therefore, at low $x$, the line connecting $F_2-\log_{10}x$ at constant $W$
would be a straight line, as a function of $\log_{10} Q^2$, 
if $F_2$ were a constant and contained no dynamics.

The parameterization
$A(x)+B(x)\log_{10}Q^2+C(x)(\log_{10}Q^2)^2$ has been used
to fit the $F_2$ measurements at each value of $x$. 
The quality of the fit is good, and
the result is shown in Figure 24 as solid lines.
The constant $W$ points on the
parameterizations have been found according to the formula
$W^2=Q^2(1/x-1)$, and are indicated on the plot with dashed lines.
It is interesting to note the distortion in the fixed-$W$ 
lines that occurs at $x \approx$10$^{-4}$ at relatively high 
$Q^2\approx$ 5 GeV$^2$ and at $W$ above 85 GeV.
\newpage

\begin{wrapfigure}{l}{4in}
\centerline{\epsfig{file=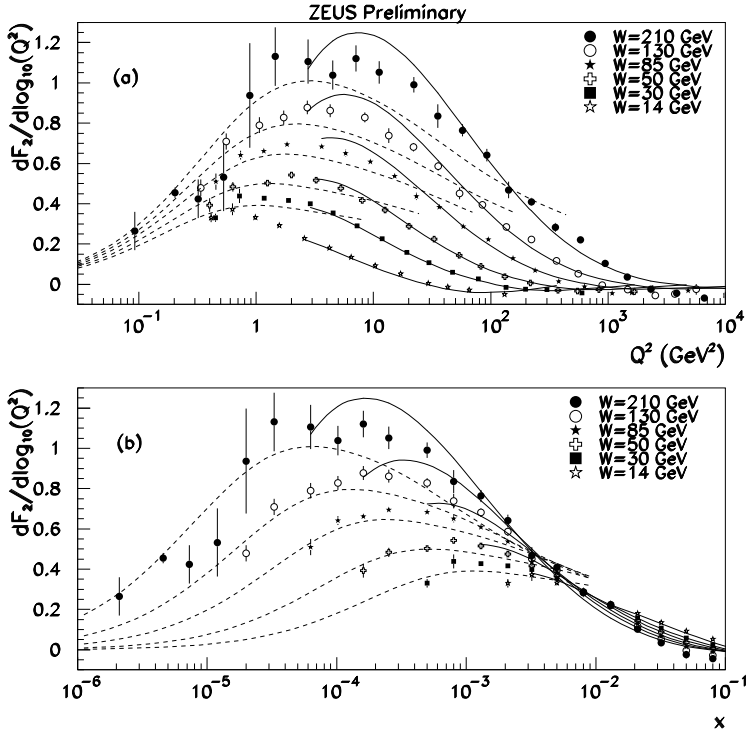,height=10cm,clip=}}
\caption{The logarithmic slope of $F_2$ at a constant $W$ as a function
of $Q^2$ and $x$.  See text.}
\end{wrapfigure}

In Figure 25, the derivatives of $F_2$, evaluated from the fit 
$B+2C\log_{10}Q^2$, are shown.  The errors of the derivatives
are evaluated using the errors on, and correlations between,
the parameters $B$ and $C$ obtained from the polynomial fits.
Figure 25a shows the derivatives at
constant $W$ as a function of $Q^2$, whereas Figure 25b shows the same 
as a function of $x$.

Figure 25a shows that at lower $Q^2$, corresponding
to lower $x$ in Figure 25b, the derivatives fall 
as $Q^2$ decreases and tend to become independent of $W$ 
at $Q^2$ of about 0.4 GeV$^2$. 
This is in line with the expectation of  Equation (11).

\begin{wrapfigure}{r}{2.5in}
\centerline{\epsfig{file=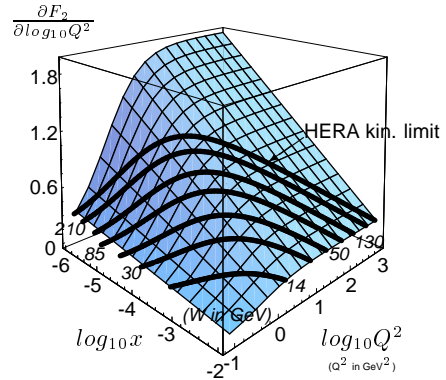,height=5.2cm,clip=}}
\caption{A schematic representation of Figure 25 in two dimensions.
See text.}
\end{wrapfigure}

Figure 25b shows that at higher $x$, corresponding
to higher $Q^2$ in Figure 25a, the derivatives fall with 
increasing $x$ and tend 
to become independent of $W$ at $x >$ 0.003, in line with the expectation
of  Equation (4), if $xg(x,Q^2)$ has the form $x^{-\lambda}$ and only
a slow dependence on $Q^2$.

The value of $Q^2$, or $x$, where 
the slope of the derivatives changes sign can be read
off from Figures 25a and 25b.  

This transition,
for values of $W$ above 85 GeV, happens at a relatively high 
$Q^2$ of 2-6 GeV$^2$ 
at the corresponding $x$ of 5$\cdot$10$^{-4}$ to 3$\cdot$10$^{-3}$.

\begin{wrapfigure}{r}{2.5in}
\centerline{\epsfig{file=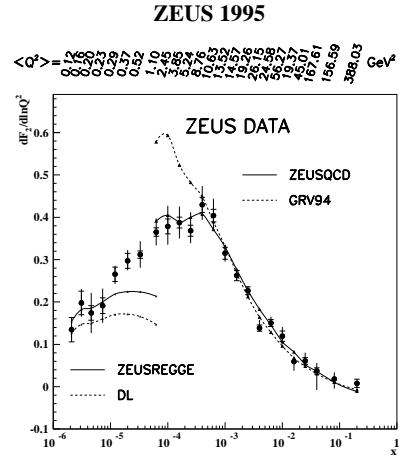,height=6cm,clip=}}
\caption{The logarithmic slope evaluated at $Q^2$ (scale at the top) and
$x$ (scale at the bottom).  See text.}
\end{wrapfigure}

The dashed lines in Figure 25 are the prediction of the GB\&W model, which 
explicitly incorporates the transition between 
the behaviors described in Equations 
(11) and (4) as a transition between the pQCD region and the
saturation region of the dipole cross-section.  
The lines are drawn only at $x$ $<$ 0.01 where
the model is applicable.  The solid line is the result of a DGLAP 
fit by the ZEUS collaboration.  The line is drawn only above 
$Q^2 > 2.7$ GeV$^2$, where data have been fit.  In case of the
DGLAP fit, the peaking behavior of Figure 25 is related to
the rapid decrease of the gluon density at low $x$ between Q$^2$ of
10 GeV$^2$ and 1 GeV$^2$ \cite{zeusdiff,mrst}.

Figure 26 is a 2-dimensional visualization of the qualitative features
of Figure 25.  The lines of constant $W$ (``trajectories"), 
plotted as thick solid lines,
when projected to the $Q^2$ and $x$ axes, approximately give 
Figures 22a and 22b,
respectively.  The peak structure of the constant $W$ lines represents
the transition from the region where $\partial F_2/\partial log_{10}x$
is steeply rising with $-log_{10}x$ to the region where it is steeply rising
with $log_{10}Q^2$, behaviors corresponding to Equations (4) and (11),
respectively.  The position of the peak structure 
does not depend strongly on the ``trajectory", $f(x,Q^2)$,
at which the derivatives 
are plotted, chosen to be $W^2 \approx Q^2/x$ in Figure 25. 
Inspection of Figure 26 shows that
almost any choice of a function $f(x,Q^2)$, 
which rises monotonically with 
decreasing $x$ for a fixed $Q^2$ will have a peak at the same $(x,Q^2)$
position which correctly indicates the transition. 
In particular, an early ZEUS analysis, made with
a more limited data set available at the time, shows (Figure 27) the
same transitional behavior at the same $(x,Q^2)$ position.

\begin{wrapfigure}{l}{4in}
\centerline{\epsfig{file=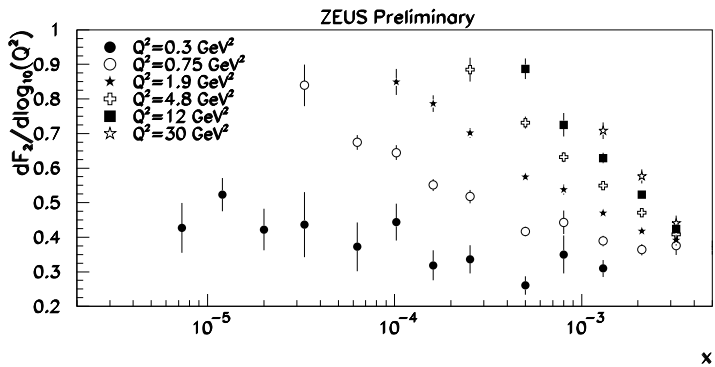,height=5cm,clip=}}
\caption{The logarithmic slope of $F_2$ for fixed $Q^2$ as a function
of $x$.}
\vspace{-0.5cm}
\end{wrapfigure}

Figure 28 shows the derivatives plotted for constant $Q^2$, as a function
of $x$.  At $Q^2$ above 1.9 GeV$^2$, 
the derivatives rise rapidly as $x$ decreases
in line with the expectations of  Equation (4) assuming
$xg(x,Q^2)$ has the form $x^{-\lambda}$.  As $Q^2$ falls, the rise
of the derivative becomes less steep, as would be expected from  Equation (11),
if a mild energy dependence, of the Regge type, is assumed for $\sigma_0$;
thus a turn-over in Figure 28 is not expected from the GB\&W model, while
naively one might expect it from  Equation (4) with saturating gluons, $xg$.
The statements sometimes found in the literature that the turn-over found in
Figure 27 is ``accidental" or ``only
kinematic in origin" \cite{mcdermott,gotsman,deroeck,schleper} arises from 
the combination of the expectation of a turn-over
in Figure 28 as a manifestation of gluon saturation  
and the misunderstanding that Figure 27
is in some way an approximation of Figure 28;
rather, it is  an approximation of Figure 25.

\section{Discussion and Outlook}

The measurements of the proton structure at small $x$ at HERA are
now very precise. However, in spite of the expectation that 
these measurements should  show some manifestation of dynamics 
beyond that incorporated in the ($\ln{Q^2}$) expansion of the DGLAP formalism,
the DGLAP fits to the data give a good description above $Q^2$ of about
1~GeV$^2$. 

The question of whether the success of DGLAP fits merely indicates
the flexibility of the parton parameterizations and the still-limited 
($\ln{Q^2}$) range of the measurement at small-$x$ is not likely to
be answered by looking at $F_2$ alone, at least in the currently 
available kinematic range.
As discussed in Section 4, while the qualitative features of
$F_2$ at small-$x$ (and necessarily small $Q^2$), show some characteristics 
expected by a saturation model, the DGLAP fits can reproduce those
same features without any parton saturation, 
as a result of a rapidly evolving gluon density at low $Q^2$.

One of the most promising ways of investigating the small-$x$ proton
structure is to look at the inclusive DIS measurements together
(``and" rather than ``or" of the title) with the diffractive DIS reactions.
While the theoretical understanding of the relation of small-$x$ and
diffraction is not yet very rigorous, the data sets from HERA
provide many interesting indications of the underlying dynamics.

\Acknowledgements
It is a pleasure to thank the organizers for a most enjoyable Symposium.
The experimental results presented here are due to a huge collaborative effort,
over many years, of the large number of people in 
the H1 and ZEUS experiments.  
I had many interesting 
discussions on the subject matter of this talk with many of 
my colleagues in H1 and ZEUS; I would particularly
like to thank Brian Foster .
I would also like to thank John Collins, Krzysztof Golec-Biernat,
Al Mueller and Mara Soares for illuminating discussions. 
Finally, Jim Whitmore, Peter Schleper and Brian Foster were kind enough
to carefully check the manuscript.

\end{document}

%% file: econfmacros.tex
\def\beq{\begin{equation}}
\def\eeq#1{\label{#1}\end{equation}}
\def\eeqn{\end{equation}}

\def\beqa{\begin{eqnarray}}
\def\eeqa#1{\label{#1}\end{eqnarray}}
\def\eeqan{\end{eqnarray}}

\let\bar=\overbar

\def\Dslash{\not{\hbox{\kern-4pt $D$}}}
\def\dslash{\not{\hbox{\kern-2pt $\del$}}}

\def\msb{{\bar{\ssstyle M \kern -1pt S}}}

